\DeclareRobustCommand\openone{\leavevmode\hbox{\small1\normalsize\kern-.33em1}}
\newcommand{\oper}[2]{\hat{#1}^{\phantom{\dag}}_{\bf #2}}
\newcommand{\operdag}[2]{\hat{#1}^{\dag}_{\bf #2}}

\documentclass[aps,prl,twocolumn,groupedaddress,showpacs]{revtex4}

\usepackage{graphicx,amsmath,amssymb,color}

\begin{document}

\title{Antiferromagnetic Order of Repulsively Interacting Fermions on Optical lattices}

\author{T. Gottwald}
\email{tobias.gottwald@uni-mainz.de}
\author{P.\ G.\ J. van Dongen}

\affiliation{KOMET 337, Institut f\"ur Physik, Johannes Gutenberg Univerit\"at, Mainz}

\date{\today}

\begin{abstract}
The N\'eel state in fermionic mixtures of two pseudospin species in an optical lattice is analyzed at low temperatures. Experimentally it remains a challenge to demonstrate antiferromagnetic correlations in ultracold fermionic quantum gases. We find that, while in balanced systems the N\'eel order parameter can point in any spatial direction, in imbalanced mixtures antiferromagnetism is strictly perpendicular to the quantization axis (i.e., the $z$-axis). Since, experimentally, one always has to assume some minimal imbalance this should have important consequences for ongoing experiments.
\end{abstract}


\maketitle

\section{Introduction}
Ultracold quantum gases on optical lattices provide an interesting experimental environment for testing elementary quantum many-body models like the Bose- or the Fermi-Hubbard model \cite{zoller:cold,bloch:many:bg9i}. In fact, microscopic quantum phenomena, predicted by theory, like correlated particle tunneling \cite{foelling:direct:bGG49} and superexchange \cite{trotzky:time:ft89J}, as well as macroscopic quantum phenomena like the Mott metal-insulator transition \cite{greiner:quantum:fgHj34,jordens:amott:gu89} have by now been observed experimentally.

\begin{figure*}
  \begin{minipage}{0.66\columnwidth}
  \resizebox{1.0\columnwidth}{!}{
  \includegraphics[clip=true]{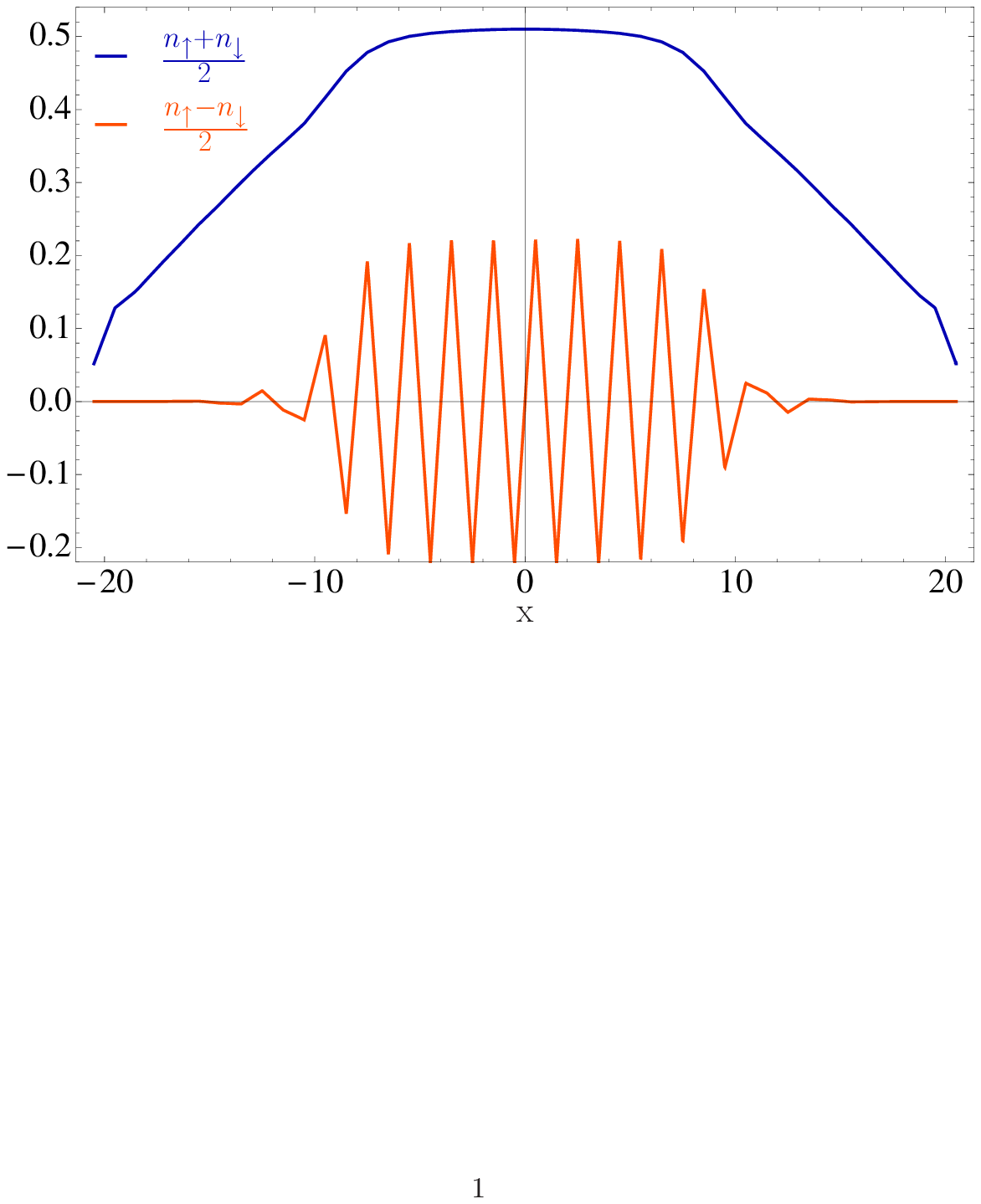}}
 \end{minipage}
 \begin{minipage}{0.66\columnwidth}
  \resizebox{1.0\columnwidth}{!}{
  \includegraphics[clip=true]{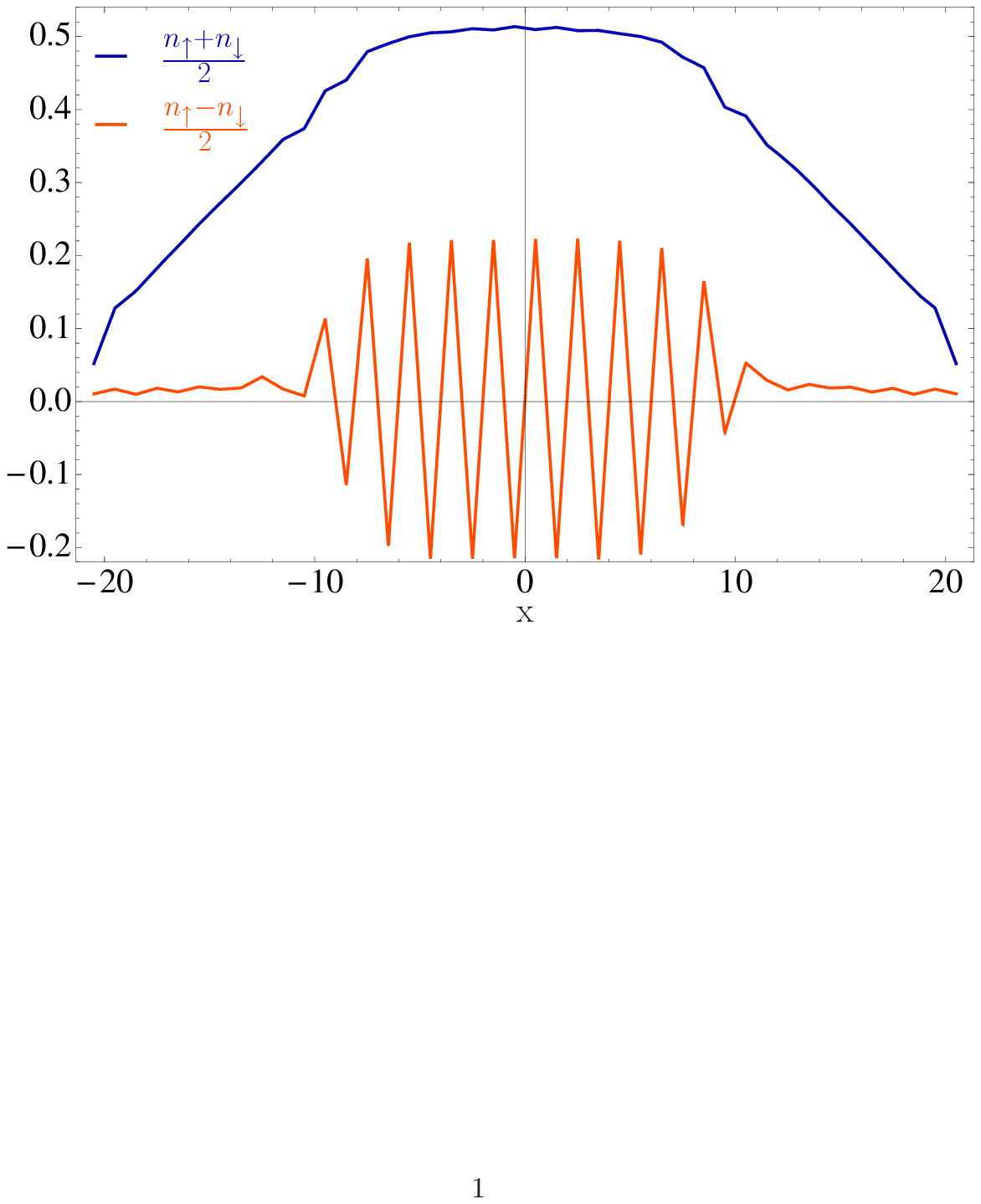}}
 \end{minipage}
 \begin{minipage}{0.66\columnwidth}
  \resizebox{1.0\columnwidth}{!}{
  \includegraphics[clip=true]{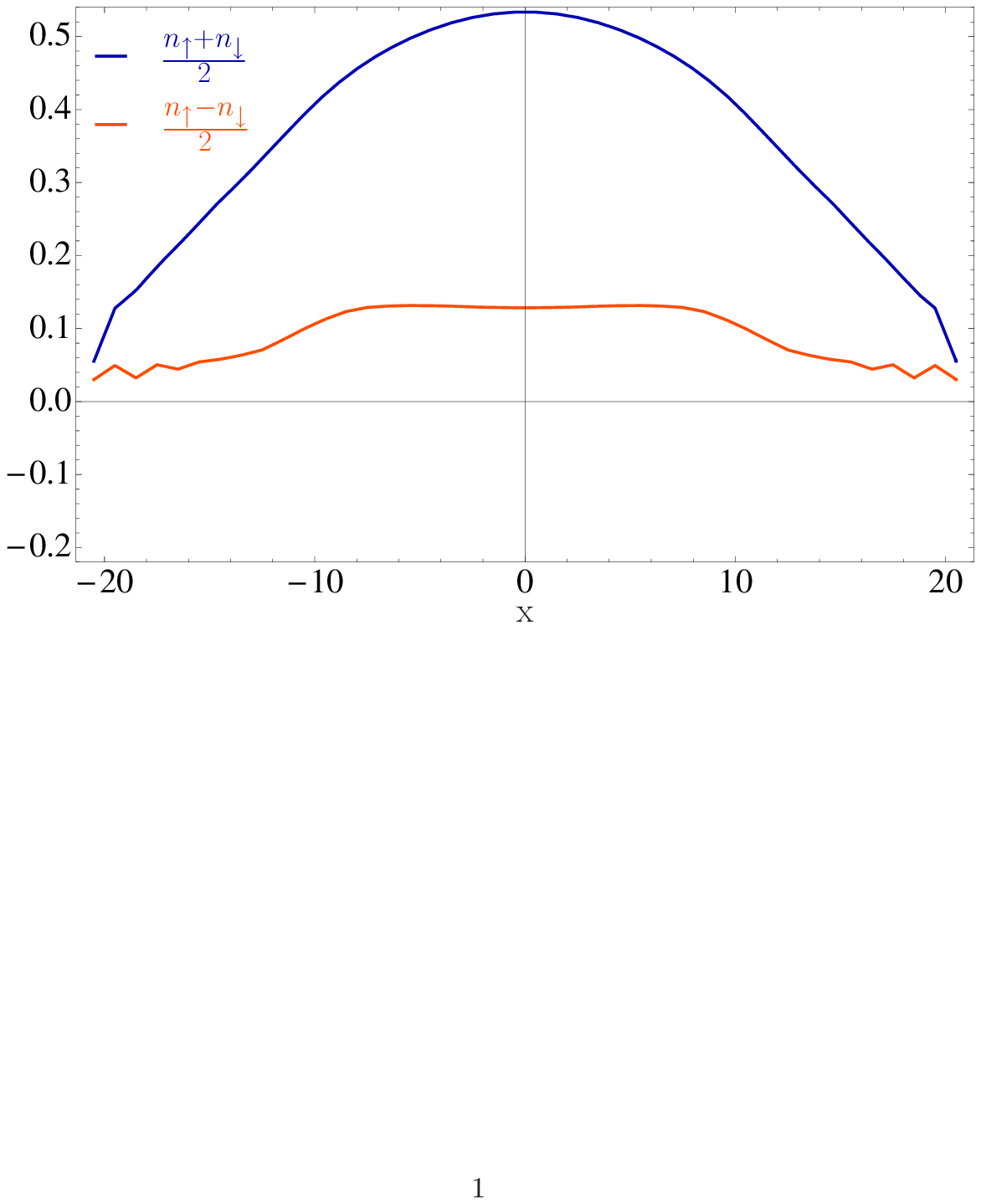}}
 \end{minipage}
 \begin{minipage}{0.66\columnwidth}
  \resizebox{1.0\columnwidth}{!}{
  \includegraphics[clip=true]{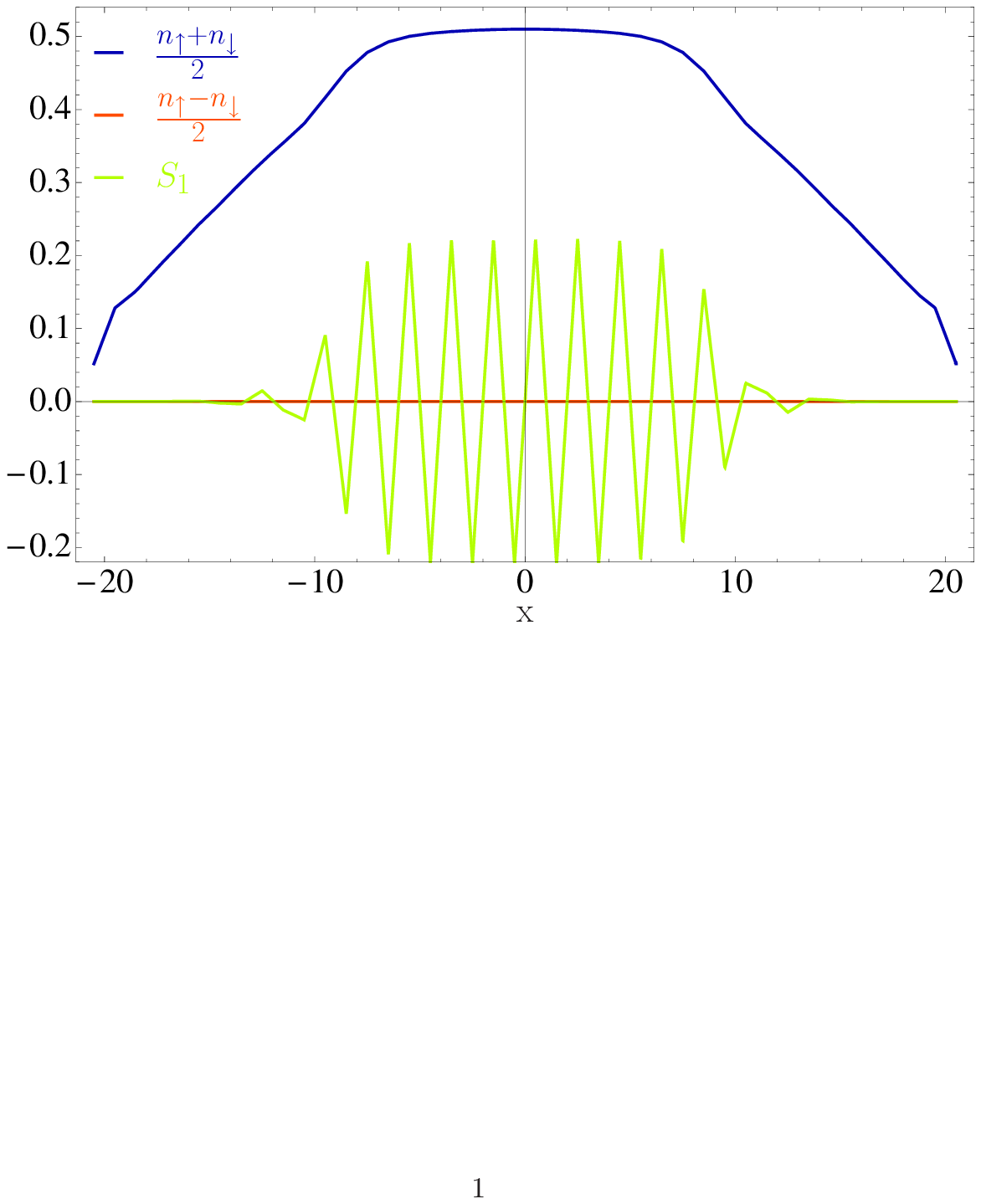}}
   \begin{center}
   (a)
   \end{center}
 \end{minipage}
 \begin{minipage}{0.66\columnwidth}
  \resizebox{1.0\columnwidth}{!}{
  \includegraphics[clip=true]{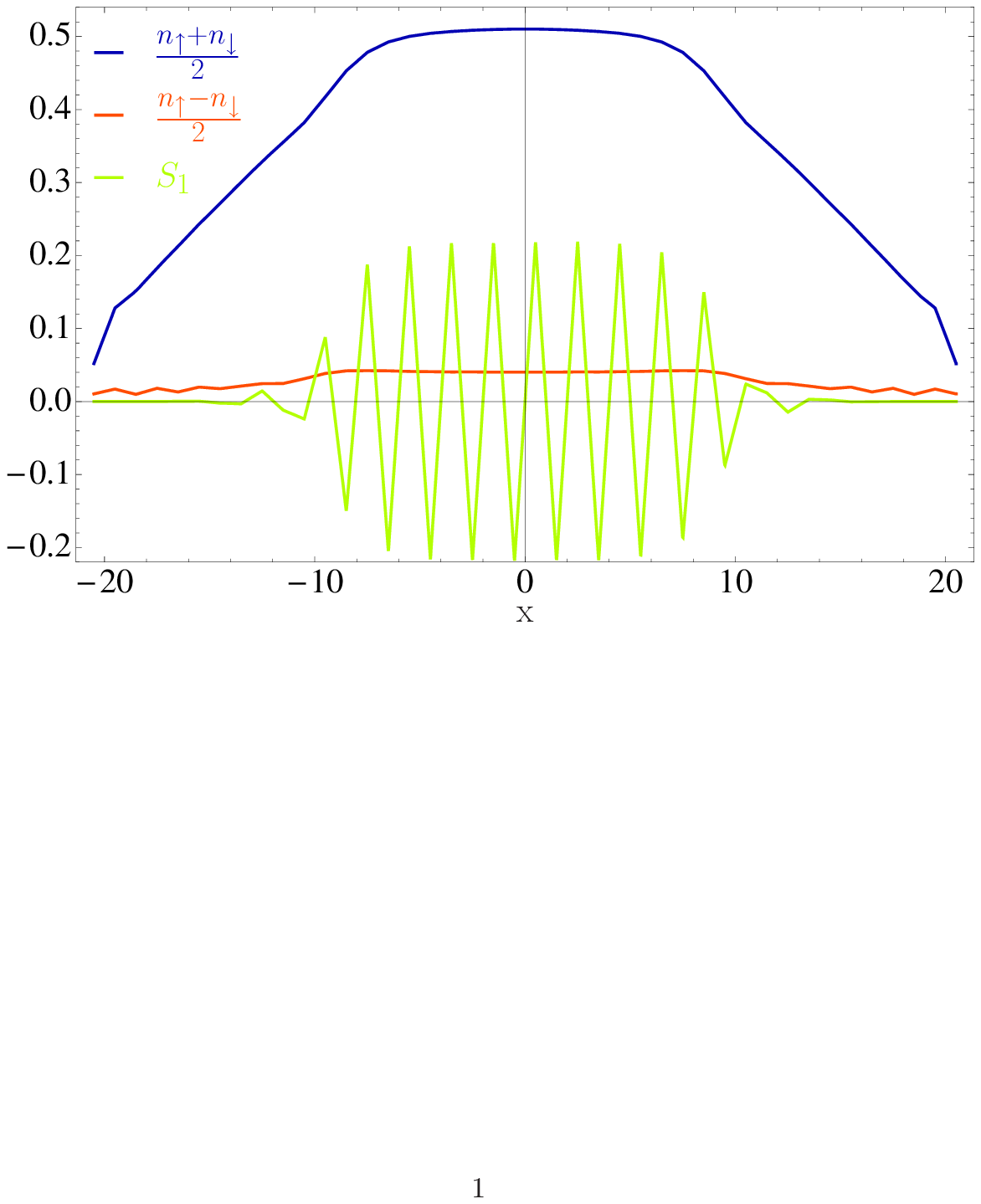}}
   \begin{center}
   (b)
   \end{center}
 \end{minipage}
 \begin{minipage}{0.66\columnwidth}
  \resizebox{1.0\columnwidth}{!}{
  \includegraphics[clip=true]{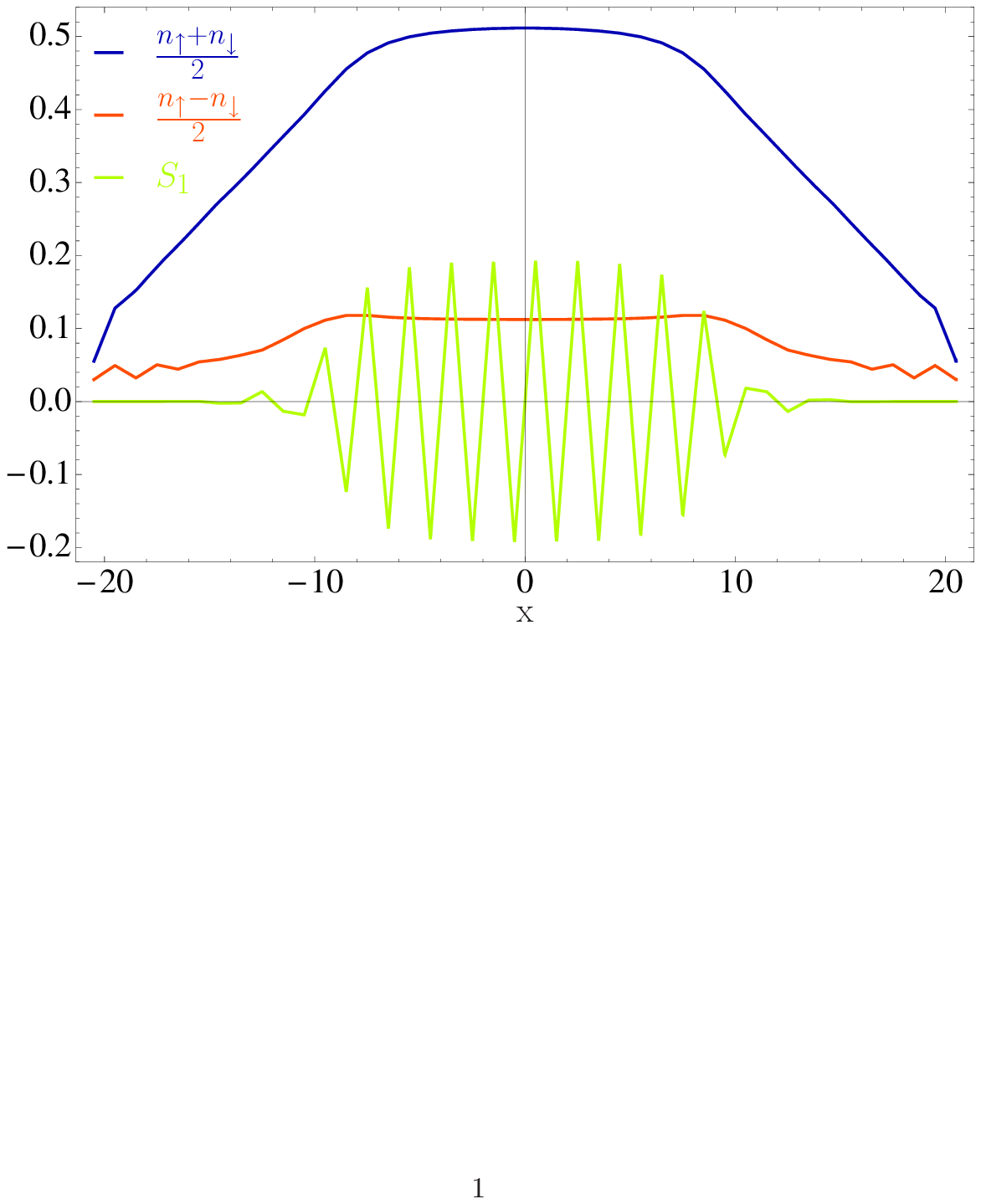}}
   \begin{center}
   (c)
   \end{center}
 \end{minipage}
 \caption{Particle density $n$ (blue/dark line) and magnetizations $\hat{S}_3$ (orange/grey line) and $\hat{S}_1$ (green/light grey line) on a $42 \times 42$ square lattice depending on $x$-position. Magnetic ordering perpendicular to the $z$-axis is suppressed in the upper panels while it is allowed in the lower panels. The parameters are chosen to be $V=0.01$, $U=2.4$, $\beta= 10$ and $\frac{\mu_\uparrow + \mu_\downarrow}{2}=1.5$ in units of $t$. The unbalance parameter $\Delta \mu = \mu_\uparrow - \mu_\downarrow$ is chosen to be (a) $\Delta \mu =0$ , (b) $\Delta \mu =0.2$ and (c) $\Delta \mu =0.6$.}\label{ut688n}
\end{figure*}

One experimental challenge for the near future is the observation of the N\'eel state in an ultracold fermionic mixture \cite{werner:interaction:5GJk7,koetsier:achieving:ko0,schneider:metallic:vfghi}. Fermi mixtures are realized with neutral fermionic atoms in different hyperfine states, for example $^{40}$K with total angular momentum $F=9/2$ in the hyperfine states $F_m = -9/2$ and $F_m = -7/2$, where the quantization axis is parallel to the magnetic field superimposed on the experimental environment \cite{zwierlein:fermionic}. It is by now experimentally feasible to detect the two different hyperfine states seperately with in-situ imaging techniques \cite{schneider:metallic:vfghi,patridge:pairing,stoof:deformation,stoof:sarma} as well as with time-of-flight methods \cite{bloch:interfrence,zwierlein:fermionic}, so that it seems a priori possible that the detection of antiferromagnetic correlations is observable with one of those pseudospin-selective detection methods \cite{bruun:probing:fgt4}. Clearly a thorough understanding of theoretical concepts and predictions will be of great help in planning any experimental investigation of the antiferromagnetic state.

In this paper we focus on the spatial structure of the N\'eel state in trapped ultracold Fermi mixtures. We show that antiferromagnetic correlations of a two spin-species Fermi mixture in the N\'eel state are expected to be spatially {\em perpendicular\/} to the $z$-axis (of the underlying Hubbard model) in general, so that pseudospin-selective measurement methods may fail to detect such transverse antiferromagnetism. In this paper we first motivate our methods, then present our results and close with a proposal on how the transverse N\'eel state may be detected experimentally.

\section{Model and Method}
A two pseudospin-species Fermi mixture on an optical lattice may be described by an inhomogeneous Hubbard model \cite{iskin:population:0FGw,andersen:magnetic:f890,snoek:antiferro:r5j}. Since, for the purposes of this paper, we are interested in the phase diagram of the Hubbard model at  low temperatures and not-too-strong repulsive interaction ($0\leq U\lesssim$ bandwidth $\approx 4t$), it suffices to study the Hubbard Hamiltonian in the saddle point approximation: 
\begin{equation}\label{chT6a}
\begin{split}
\mathcal{H} = -t \sum_{( {\bf ij} ), \sigma} \operdag{c}{i \sigma} \oper{c}{j \sigma} + \sum_{\bf i \sigma} \left( V{\bf i}^2 - \mu_{\sigma} \right) \oper{n}{i \sigma} \\
+ U \sum_{\bf i} \left( 2 \langle \oper{n}{\bf i} \rangle \oper{n}{\bf i} - 2 \langle \oper{\bf S}{\bf i} \rangle\! \cdot\! \oper{\bf S}{\bf i} - \langle \oper{n}{\bf i} \rangle^2  +  \langle \oper{\bf S}{\bf i} \rangle^2 \; \right),
\end{split}
\end{equation}
where $\oper{c}{i \sigma}$ is the fermionic annihilation operator for a fermion at site ${\bf i}$ with pseudospin $\sigma$ (e.g., $\uparrow \hat{=}$ $F_m = -9/2$ and $\downarrow \hat{=}$ $F_m = -7/2$), $\oper{n}{i \sigma} = \operdag{c}{i \sigma} \oper{c}{i \sigma}$, $\oper{n}{\bf i}= \frac{1}{2} ( \oper{n}{\bf i \uparrow} +\oper{n}{\bf i \downarrow} ) $ and  $\oper{\bf S}{\bf i} $ is the local spin operator. Furthermore, $t$ is the nearest-neighbor hopping amplitude, $V$ is the confining potential strength and $U$ describes the on-site repulsion. The spin-dependent chemical potential $\mu_\sigma$ controls the particle numbers in the grand-canonical formalism. The saddle point approximation \eqref{chT6a} of the Hubbard Hamiltonian is able to describe antiferromagnetic correlations both in $z$-direction and in the $xy$-plane. It is expected to yield qualitatively correct results at not-too-large coupling ($U \lesssim $ bandwidth) but is known to quantitatively overestimate energy scales (gaps, critical temperatures) \cite{dongen:gt8Yaqv4}. However, it is difficult to determine the accuracy of the saddle-point approximation exactly for a trapped inhomogeneous systems. Since the filling is space-dependent and the system has a finite size one would not expect the logarithmic van Hove singularity in the non-interacting density of states in two dimensions to cause a divergent contribution in the second order perturbation expansion \cite{dongen:gt8Yaqv4}. Hence, the results for two dimensions are expected to be similar to the ones in three dimensions and therefore it is our best guess is that the order parameter in the saddle-point approximation is suppressed by a factor of two to four by quantum fluctuations.

For balanced systems, $\Delta \mu \equiv \mu_\uparrow - \mu_\downarrow=0$, the Hamiltonian \eqref{chT6a} is fully invariant under spin-rotations of the form
\begin{equation}
 \left(
 \begin{array}{c}
  \oper{c}{i \uparrow} \\
  \oper{c}{i \downarrow}
 \end{array}
 \right)
 \; \rightarrow\;
 \mathcal{U}
 \left(
 \begin{array}{c}
  \oper{c}{i \uparrow} \\
  \oper{c}{i \downarrow}
 \end{array}
 \right) 
 \quad , \quad
 \mathcal{U} \in { SU(2)}
 \; ,
\end{equation}
while the symmetry group reduces to $U(1)$ for any $\Delta \mu \neq 0$. Since one expects the grand potential of any self-consistent antiferromagnetic solution to have the same global symmetry as the Hamiltonian \eqref{chT6a}, the N\'{e}el state should break a $SU(2)$-symmetry for $\Delta \mu = 0$ and an $U(1)$-symmetry for $\Delta \mu \neq 0$.

\begin{figure*}[ht]
  \begin{minipage}{0.66\columnwidth}
  \resizebox{1.0\columnwidth}{!}{
  \includegraphics[clip=true]{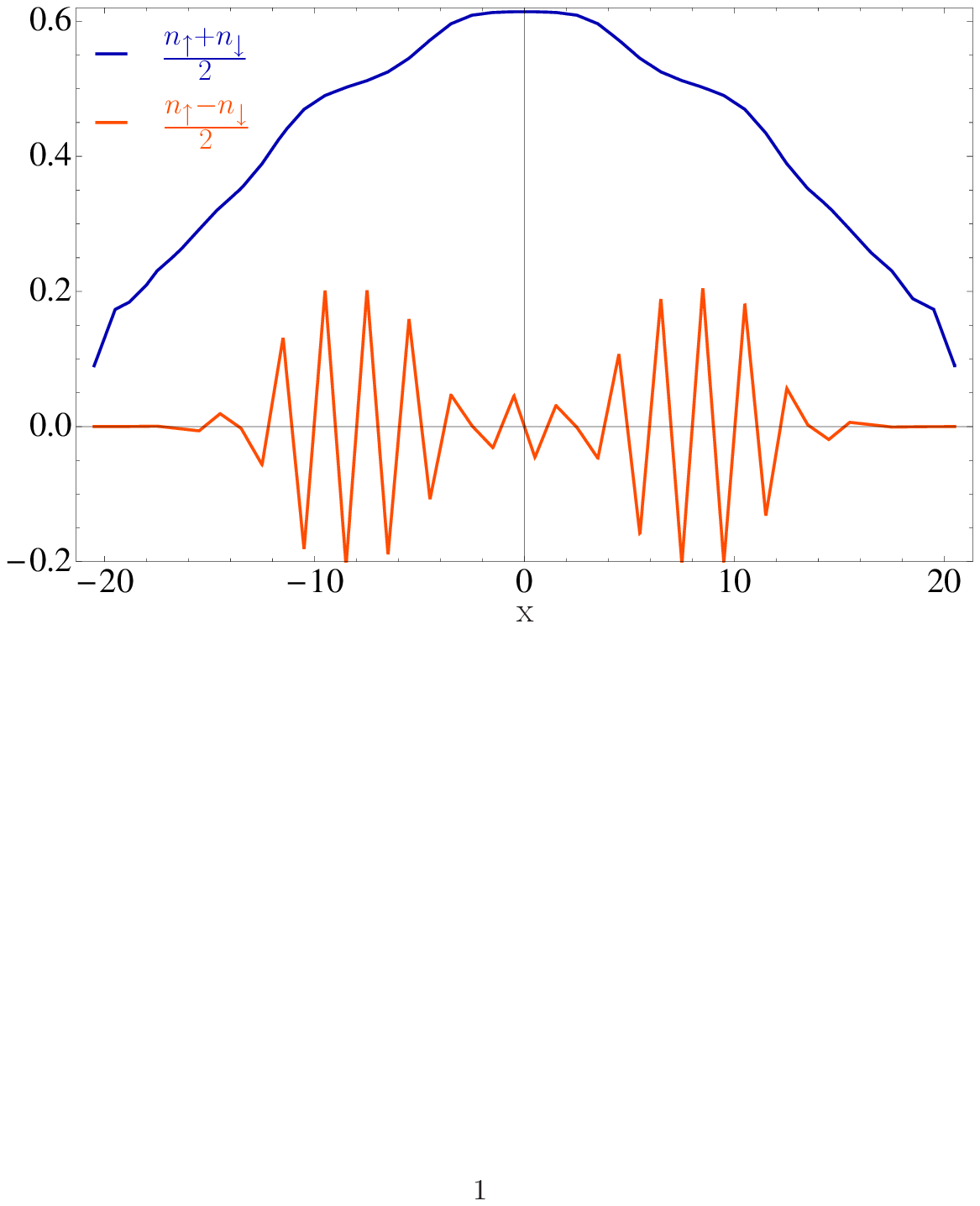}}
 \end{minipage}
 \begin{minipage}{0.66\columnwidth}
  \resizebox{1.0\columnwidth}{!}{
  \includegraphics[clip=true]{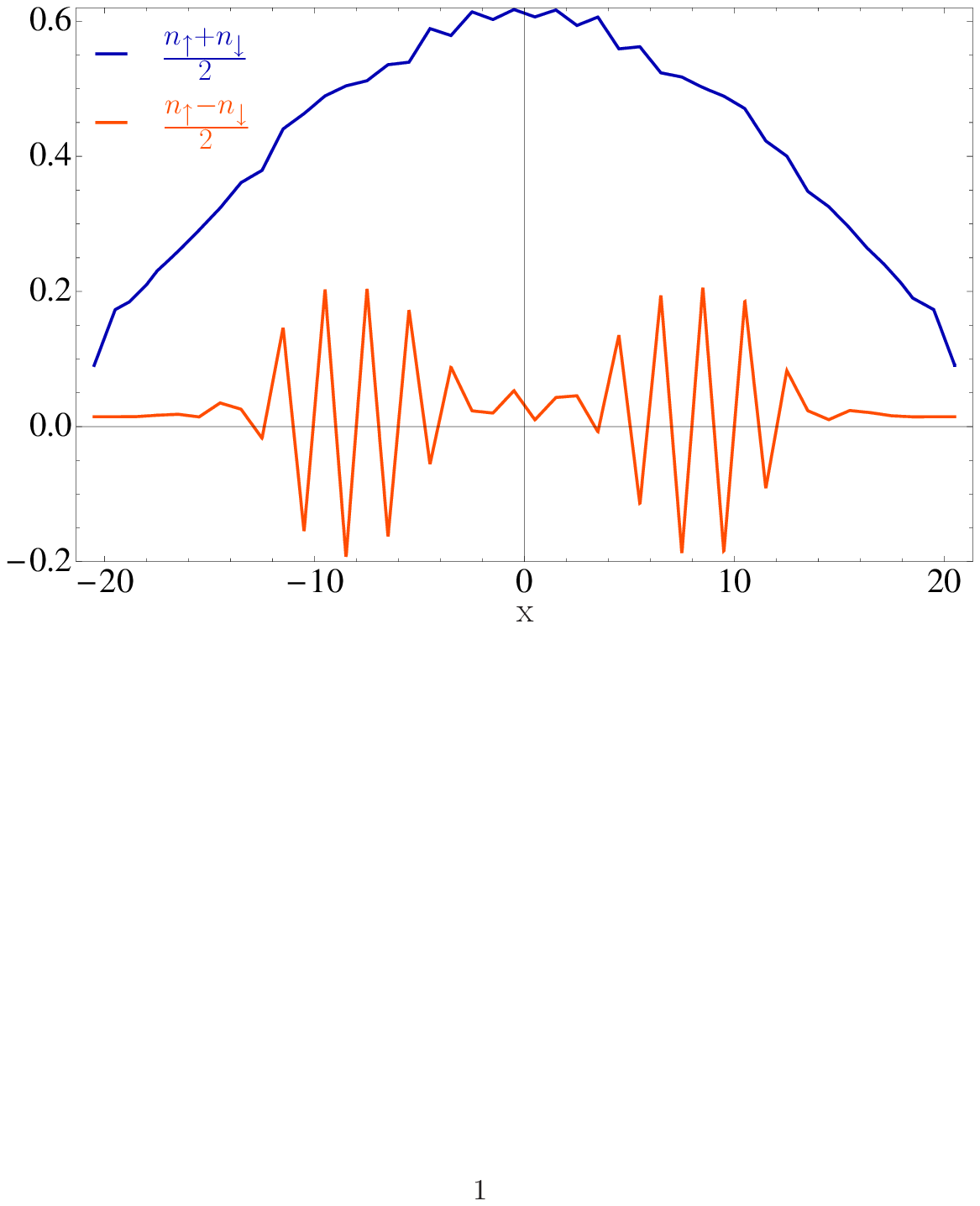}}
 \end{minipage}
 \begin{minipage}{0.66\columnwidth}
  \resizebox{1.0\columnwidth}{!}{
  \includegraphics[clip=true]{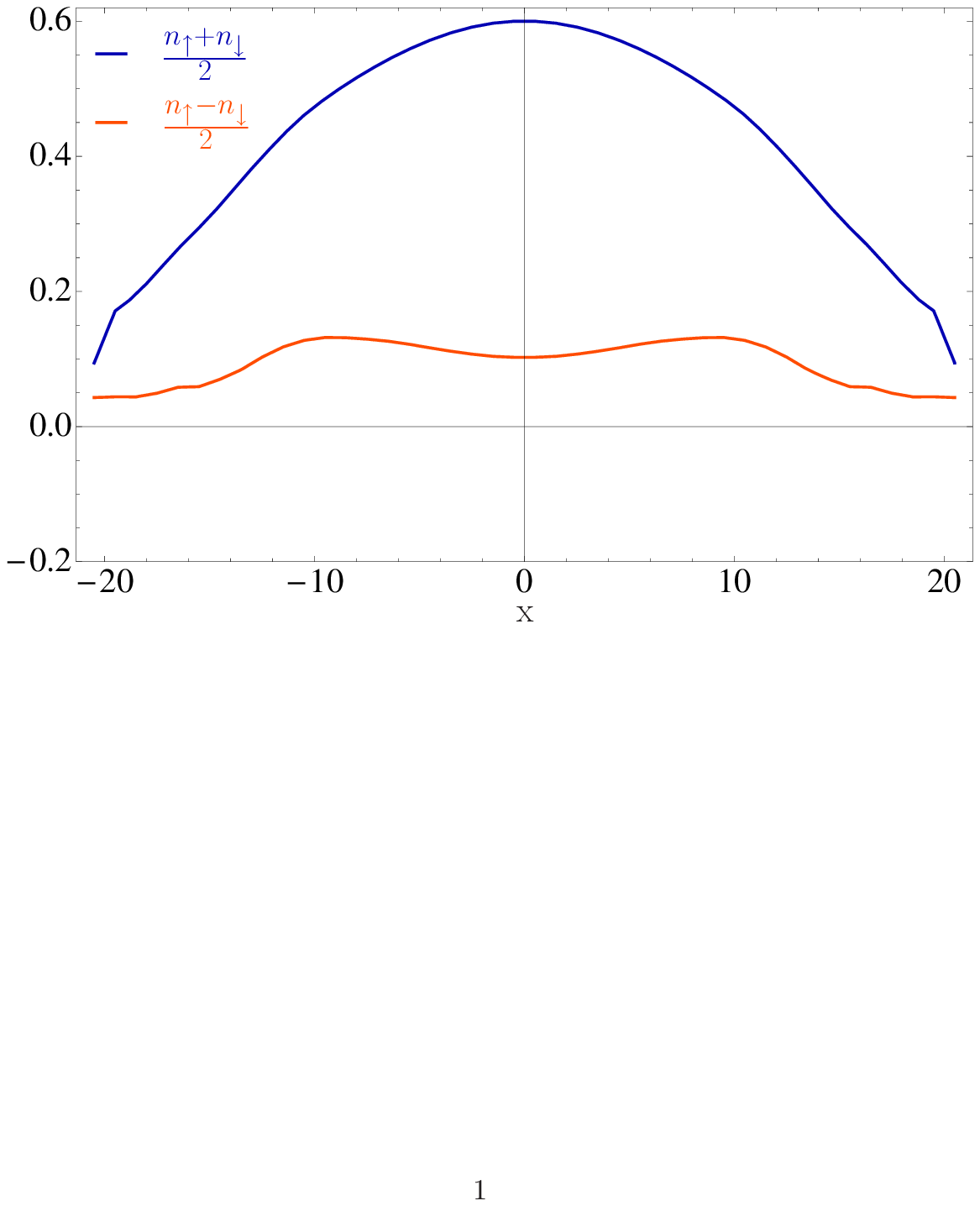}}
 \end{minipage}
 \begin{minipage}{0.66\columnwidth}
  \resizebox{1.0\columnwidth}{!}{
  \includegraphics[clip=true]{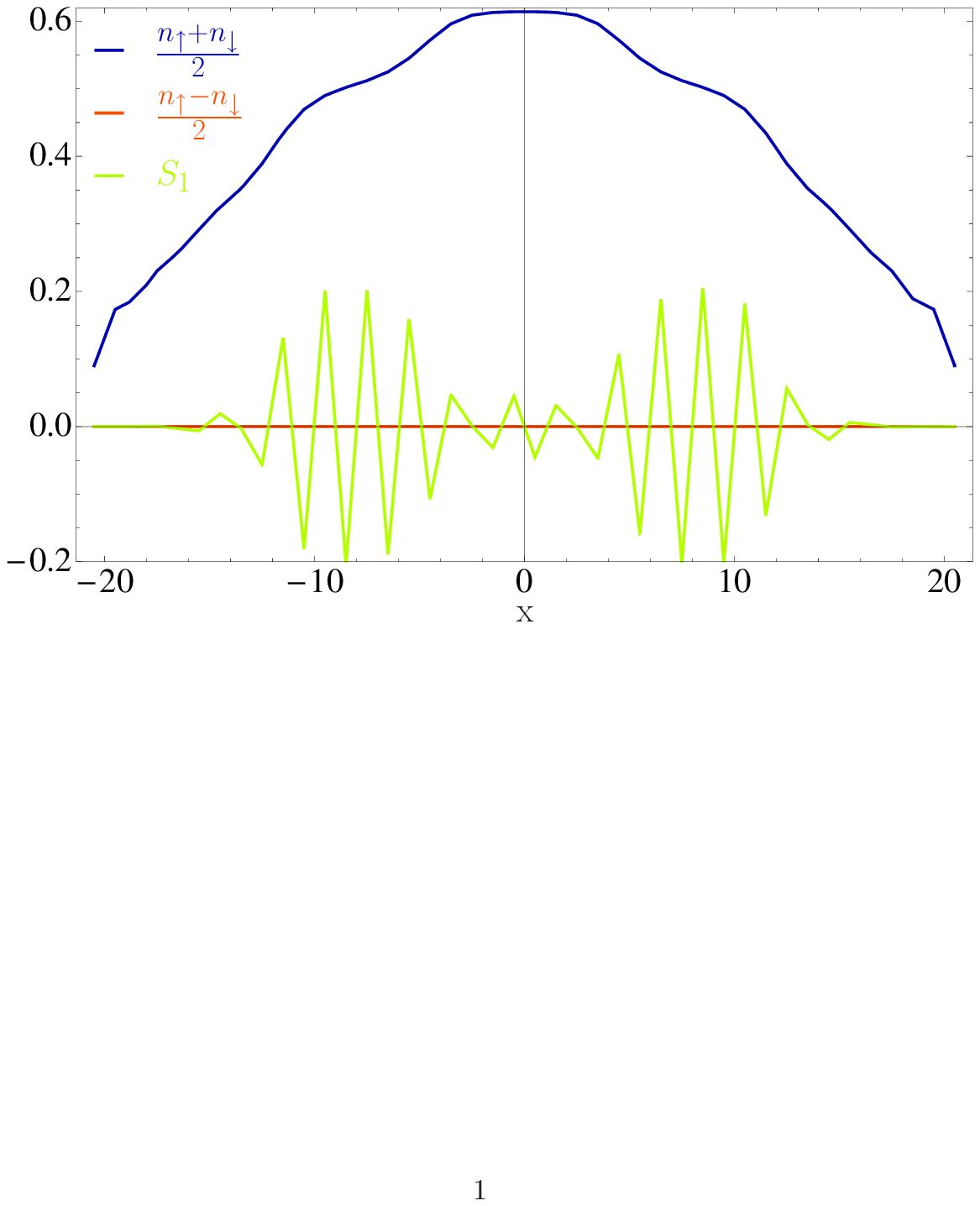}}
   \begin{center}
   (a)
   \end{center}
 \end{minipage}
 \begin{minipage}{0.66\columnwidth}
  \resizebox{1.0\columnwidth}{!}{
  \includegraphics[clip=true]{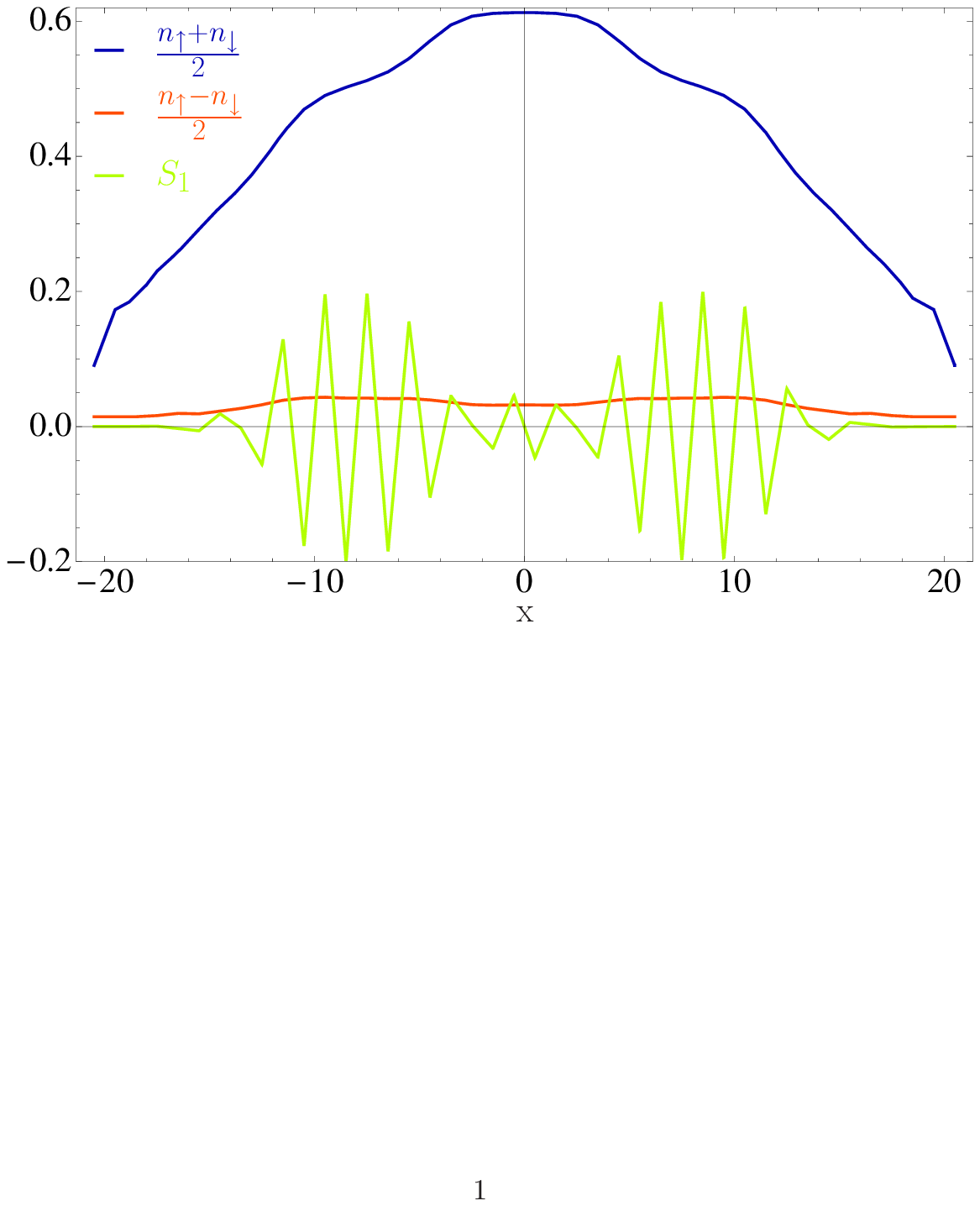}}
   \begin{center}
   (b)
   \end{center}
 \end{minipage}
 \begin{minipage}{0.66\columnwidth}
  \resizebox{1.0\columnwidth}{!}{
  \includegraphics[clip=true]{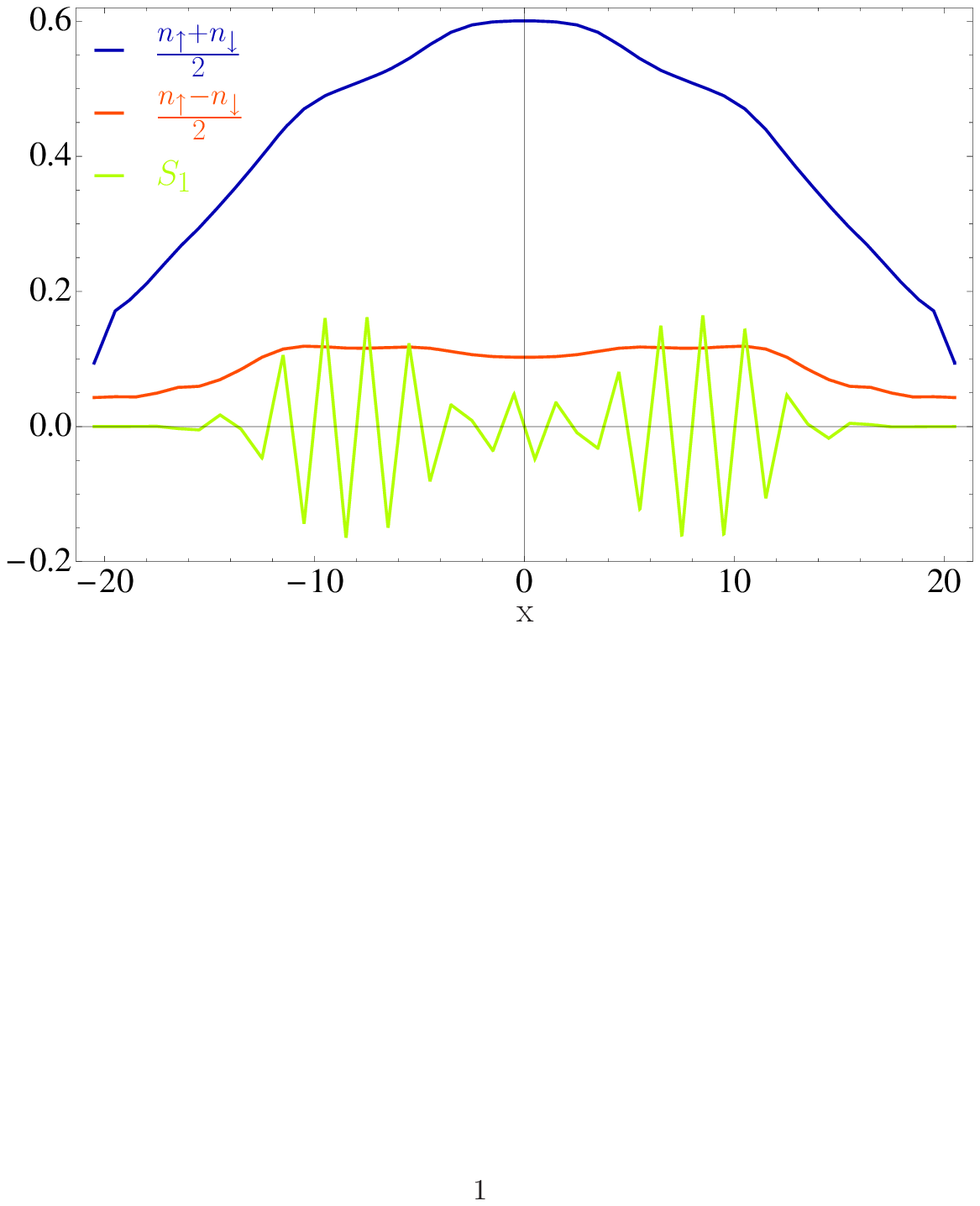}}
   \begin{center}
   (c)
   \end{center}
 \end{minipage}
 \caption{Particle density $n$ (blue/dark line) and magnetizations $\hat{S}_3$ (orange/grey line) and $\hat{S}_1$ (green/light grey line) on a $42 \times 42$ square lattice depending on $x$-position. Magnetic ordering perpendicular to the $z$-axis is suppressed in the upper panels while it is allowed in the lower panels. The parameters are chosen to be $V=0.01$, $U=2.4$, $\beta= 10$ and $\frac{\mu_\uparrow + \mu_\downarrow}{2}=2.0$ in units of $t$. The unbalance parameter $\Delta \mu = \mu_\uparrow - \mu_\downarrow$ is chosen to be (a) $\Delta \mu =0$ , (b) $\Delta \mu =0.2$ and (c) $\Delta \mu =0.6$.}\label{gHju9}
\end{figure*}

The Hamiltonian \eqref{chT6a} can conveniently be written in the following matrix form:
\begin{equation}\label{lpwert}
\mathcal{H} = \sum_{\bf i,j}
\left(
\begin{array}{c}
 \operdag{c}{j \uparrow} \\
 \operdag{c}{j \downarrow}
\end{array}
\right)^{\!\rm T}\!\! 
\left( 
\begin{array}{cc}
 \mathcal{H}^{\phantom{\dag}}_{0 \uparrow} &  U \langle \oper{S}{-} \rangle \\
 U \langle \oper{S}{+} \rangle & \mathcal{H}^{\phantom{\dag}}_{0 \downarrow}
\end{array}
\right)_{\bf ij} \!\! 
\left(
\begin{array}{c}
 \oper{c}{i \uparrow} \\
 \oper{c}{i \downarrow}
\end{array}
\right) \; + {\rm const} \; ,
\end{equation}
where we defined $\oper{S}{\pm} = \oper{S}{\rm 1} \pm i \oper{S}{\rm 2}$, and $\mathcal{H}_{0 \sigma}$ contains the spin-diagonal terms of the Hamiltonian \eqref{chT6a}. The self-consistency equations take the form:
\begin{eqnarray}
 \label{fg8UUk}
 \langle \oper{n}{i \sigma} \rangle &=& \sum_{s=1}^{2 \mathcal{N}}  |u_{\textbf{i}s, \sigma}|^2 f_{\beta}(E_{s})  \\
 \label{pasE3}
 \langle \oper{S}{i +} \rangle &=& \sum_{s=1}^{2 \mathcal{N}} u_{\textbf{i}s,\uparrow}^* u_{\textbf{i}s,\downarrow}^{\vphantom *} f_{\beta}(E_{s}) \; ,
\end{eqnarray}
where $u_{\textbf{i}s,\sigma}$ are the components of an unitary transformation diagonalizing \eqref{lpwert}, $E_s$ are the eigenvalues of the saddle point Hamiltonian \eqref{chT6a}, $\mathcal{N}$ is the total number of lattice sites and $f_{\beta}(x) \equiv [1+\exp{(\beta x)}]^{-1}$ is the Fermi function at inverse temperature $\beta \equiv 1/k_{\rm B}T$.

We determine antiferromagnetic solutions self-con\-sis\-tently by iterating the self-consistency equations \eqref{fg8UUk} and \eqref{pasE3} in combination with the Hamiltonian \eqref{lpwert}. In order to find solutions with restricted antiferromagnetic order (purely in $z$-direction) we start the iteration cycle with $\langle \oper{\bf S}{ {\rm 1/2}, \bf i} \rangle = 0$ for all lattice sites ${\bf i}$ in the initial state. This automatically brings the Hamiltonian \eqref{lpwert} into a block-diagonal structure and, therefore, no antiferromagnetism in $xy$-direction occurs at any iteration step, since in this case at least one of the coefficients $u_{{\bf i}s,\uparrow}$ or $u_{{\bf i}s,\downarrow}$ vanishes for all indices $\{\textbf{i},s\}$. We also solve the more general problem with possibly also $\langle \oper{\bf S}{{\rm 1/2, \bf i}} \rangle \neq 0$ and compare the grand potentials of both solutions at the same values of the parameters $V$, $t$, $U$, $\mu_\uparrow$, $\mu_\downarrow$ and $\beta$.

\section{Results}
We now present a selection of numerical results for different spin-averaged chemical potentials $\mu \equiv \frac{1}{2}(\mu_\uparrow + \mu_\downarrow)$ and different imbalance strengths on a $42 \times 42$ square lattice at fixed parameters $V=0.01$, $U=2.4$ and $\beta =10$ in units of the nearest-neighbor hopping amplitude $t$. In Figure \ref{ut688n} we choose $\mu=1.5$ so that an approximately half-filled region (i.e., with one particle per site) lies in the center of the trap. In Figure \ref{gHju9} we increase the filling and choose $\mu=2.0$, so that the center of the trap  is more than half-filled. Our results also confirm the suggestion of \cite{andersen:magnetic:f890,snoek:antiferro:r5j}, that antiferromagnetic correlations are enhanced in regions which are nearly half-filled.

In general, we may obtain solutions with antiferromagnetic correlations in $z$-direction as well as in the $xy$-plane. Results for a system with a balanced number of particles are shown in Figures \ref{ut688n}(a) and \ref{gHju9}(a). The spin-rotational symmetry of the Hamiltonian \eqref{chT6a} at $\mu_\uparrow=\mu_\downarrow$ is not broken, and, accordingly, the amplitudes of the magnetization components $\langle\oper{S}{\rm 3}\rangle$ (upper panel, assuming $\langle\oper{S}{\rm 1}\rangle=0$) and $\langle\oper{S}{\rm 1}\rangle$ (lower panel, choosing $\langle\oper{S}{\rm 3}\rangle=0$) are identical in Figures \ref{ut688n}(a) and \ref{gHju9}(a). The grand potentials of both solutions are, of course, degenerate.

In unbalanced systems ($\mu_\uparrow \neq \mu_\downarrow$) the spin-rotational symmetry is reduced from $SU(2)$ to $U(1)$ and the Hamiltonian \eqref{chT6a} is invariant only under rotations around the $z$-axis. Results with the magnetization restricted to the $z$-direction are shown in the upper panels of Figures \ref{ut688n} and \ref{gHju9}, and results allowing also for $\langle\oper{S}{\rm 1}\rangle\neq 0$ are shown in the lower panels. As one can see from Figures \ref{ut688n} and \ref{gHju9}, antiferromagnetic order decreases as the imbalance (i.e., $\Delta \mu$) is increased. In Figures \ref{ut688n}(b) and \ref{gHju9}(b) numerically stable solutions are presented with antiferromagnetic order in the $z$- and the $x$-direction, respectively. Note that the solution with its order parameter perpendicular to the $z$-axis has a lower grand potential and is therefore the thermodynamically stable one. In Figures \ref{ut688n}(c) and \ref{gHju9}(c) results with a still stronger imbalance are shown. While antiferromagnetism purely in $z$-direction is not numerically stable in this case, staggered order still persists in the $xy$-plane and lowers the grand potential.

We exclusively found solutions with $S_{+, \bf i}\exp{(-i \phi)} \in \mathbb{R}$ and a site-{\em independent\/} phase $\phi$, so that the antiferromagnetic order parameter in the $xy$-plane can by a canonical transformation always be chosen to be perpendicular to the $y$-axis. This result is a priori not at all obvious, but none of our numerical investigations yielded a self-consistent solution with a more complicated structure in the $xy$-plane. In the following we assume that the spins have been globally rotated such that $S_{2, \bf i} =0$ for all ${\bf i}$.

\begin{figure}[!t]
 \begin{center}
  \resizebox{0.8\columnwidth}{!}{
  \includegraphics[clip=true]{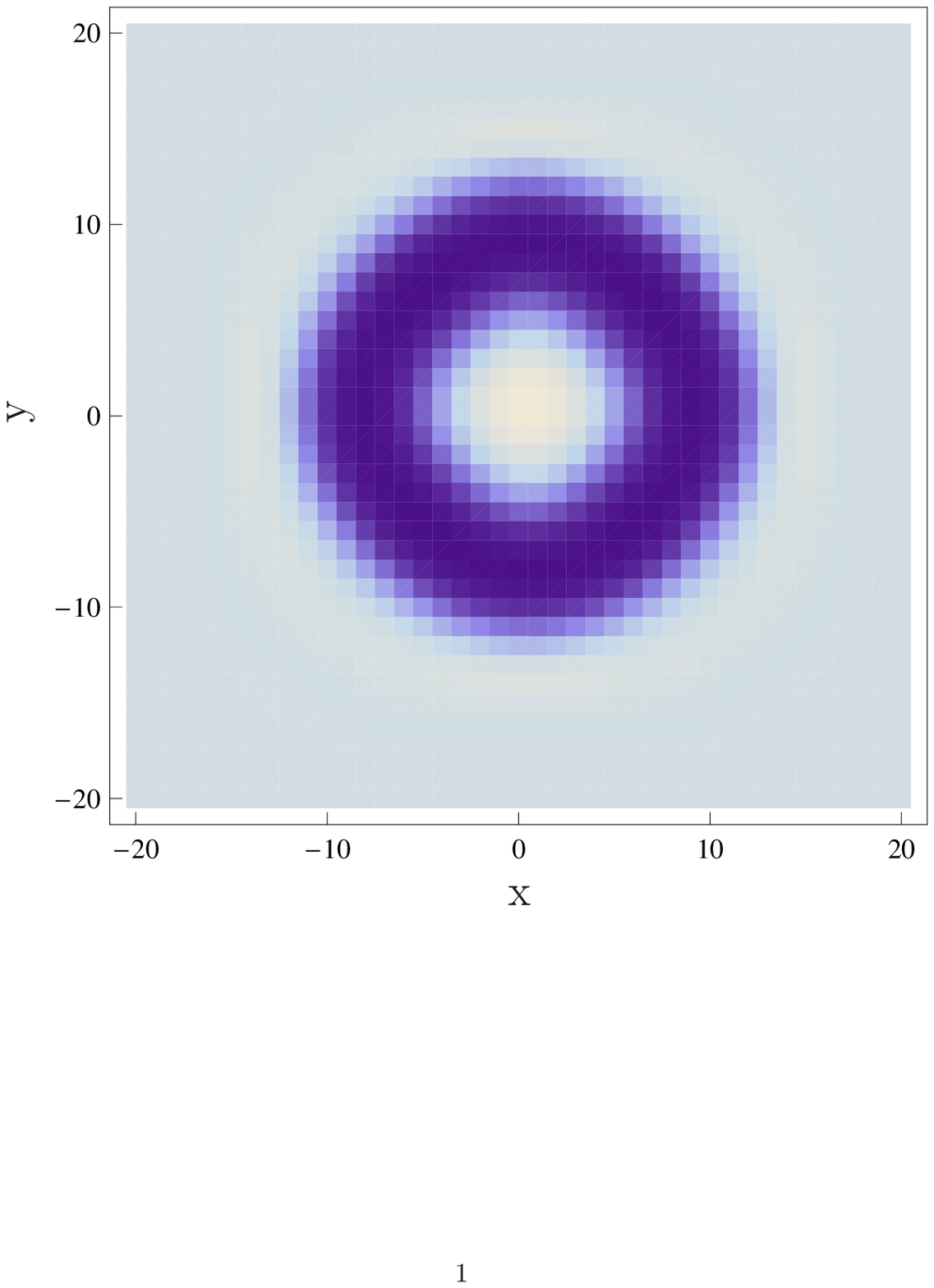}}
 \end{center}
 \caption{Staggered antiferromagnetic order parameter $m \equiv {\bf S}_{1, \bf i}  \exp (i \textbf{Q} \cdot \textbf{i})$, with $\textbf{Q}= (\pi,\pi)^{\rm T}$, on a $42 \times 42$ square lattice. The sign of $m$ is positive (negative) in the dark (light) regions, indicating the formation of different antiferromagnetic domains, separated by walls of parallelly aligned spins. In particular, the sign of $m$ is positive in the region of approximate half filling ($n_{\bf i} \approx 0.5$) and changes where the gradient of the filling factor is maximal. The parameters are chosen to be $V=0.01$, $U=2.4$, $\beta= 10$, $\Delta \mu =0$ and $\frac{\mu_\uparrow + \mu_\downarrow}{2}=2.0$ in units of $t$. The sign changes are also characteristical for unbalanced systems ($\Delta \mu \neq 0$).}\label{fuu9sw1}
\end{figure}

Incommensurate states in the translationally invariant Hubbard model have been postulated in the literature already long ago \cite{schulz:incommensurate:bg89I}. Here we find a somewhat different result, with a {\em checkerboard structure\/} for the  $S_{3, \bf i}$- and $S_{1, \bf i}$-components of the antiferromagnetic order parameter in several spatially distinct regions, which are separated by {\em domain walls\/}. These domain walls occur in regions where the total filling changes from $n_{\bf i} \equiv \langle \hat{n}_{\bf i} \rangle \approx 0.5$ to smaller or bigger values. We illustrate this domain wall formation in Figure \ref{fuu9sw1}, where we plot $S_{1, \bf i} \exp (i \textbf{Q} \cdot \textbf{i})$, with $\textbf{Q}\equiv (\pi,\pi)^{\rm T}$, as a function of the spatial coordinates $x$ and $y$. A perfectly commensurate N\'eel state in the balanced homogeneous Hubbard model would not have any sign change. In contrast, we find sign changes in every thermodynamically stable self-consistent antiferromagnetic pattern in either balanced ($\Delta \mu =0$) or imbalanced ($\Delta \mu \neq 0$) systems. Furthermore, we find that any solution minimizing the grand potential systematically has a antisymmetrical spatial structure in quantities describing antiferromagnetic correlations (e.g., $S_{1, \bf i}$ at any $\Delta \mu$-value or $S_{3, \bf i}$ at $\Delta \mu =0$), while the other quantities  (e.g., $n_{\bf i}$ and  $S_{3, \bf i}$ at $\Delta \mu \neq 0$) have a spatially symmetrical structure.

\section{Experimental consequences} We expect spin-se\-lec\-tive measurement methods like in-situ imaging or time-of-flight (TOF) techniques to be {\em inadequate\/} \cite{bruun:probing:fgt4} for the successful detection of the N\'eel state in ultracold Fermi mixtures on two-dimensional optical lattices, since the antiferromagnetic order parameter is predicted to be {\em perpendicular\/} to the $z$-axis at any finite imbalance. While the experimental techniques mentioned above are able to distinguish between two hyperfine states (e.g., $F_m = -9/2$ and $F_m = -7/2$), which corresponds to the measurement of $S_3=\frac{1}{2}(n_\uparrow - n_\downarrow)$, they are not able to capture coherent linear combinations represented by anomalous expectation values like $S_1 = (S_+ + S_-)/2$. 

Recently a method for reconstructing the non-diagonal elements of a real-space Green function through a combination of a double pseudospin-mixing Raman pulse and a TOF measurement was 
proposed \cite{duan:detecting:p0da}. However, this proposal for detecting transverse components of the order parameter, too, seems problematic due to the remaining $U(1)$-symmetry of the antiferromagnetic state. As a 
consequence of this symmetry, the direction of the transverse AFM order parameter (and, hence, the non-diagonal elements of the Green function) will be different in each TOF image, effectively preventing or at least seriously complicating the reconstruction of the Green function.

We, therefore, propose to follow a different strategy: An observable that is readily accessible using in-situ imaging techniques \cite{patridge:pairing,stoof:deformation,stoof:sarma,schneider:metallic:vfghi} is the spatial distribution of particle numbers $n_{\bf i}$. Our calculations show that antiferromagnetism leads to a {\em broadening\/} of the approximately half-filled region ($n_{\bf i} \approx 0.5$), even if the interaction is weak. This signature can be seen very clearly in Figures \ref{ut688n} and \ref{gHju9}, which show that the antiferromagnetic states have a {\em plateau\/} in the half-filled region (``wedding-cake structure''), that is characteristically absent in the paramagnetic state [upper plot of the (c)-series]. Our proposal is to use this plateau in the density profile as a smoking gun for antiferromagnetism. In doing the experiment, it will of course be crucial to choose the parameters of the experimental setup such that the plateau becomes visible within the resolution of in-situ imaging. Since regions of sizes down to approximatively $5 \times 5$ lattice sites are resolvable, parameters similar to the ones chosen in Figure \ref{ut688n} seem to be adequate for this purpose. Obviously, a plateau at half-filling might also arise at strong interaction (for $U>U_{\rm c} \gtrsim $  bandwidth) due to a Mott metal-insulator transition within a paramagnetic state \cite{jordens:amott:gu89,bloch:formation,helmes:bf5sc8H}. Hence, it would not be possible to distinguish clearly between paramagnetic Mott-in\-sulating states and antiferromagnetic ones in the strong-interaction regime. Our proposal is, therefore, to try to detect the broadening of the approximately half-filled region as a signature for antiferromagnetic correlations in the {\em weak-coupling} regime ($U<U_{\rm c}$), where the plateau cannot be caused by a Mott transition. This strategy should be practicable, since the interaction strength $U$ in ultracold quantum gases is straightforwardly tunable experimentally \cite{bloch:many:bg9i}.

\section{Summary and Conclusions} We investigated a two pseudospin-species Hubbard-model well suitable to describe spatially inhomogeneous N\'eel states of Fermi mixtures on a two-dimensional optical lattice in a trap. In our saddle point approximation we found the antiferromagnetic order parameter to have an arbitrary direction in a balanced mixture [in accordance with the full $SU(2)$-symmetry of the Hamiltonian] and to be strictly perpendicular to the quantization axis at any imbalance strength [in accordance with the $U(1)$-symmetry for $\Delta \mu \neq 0$]. Experimentally, one always has to assume some minimal imbalance. In order to experimentally observe these staggered magnetic structures, we propose to use in-situ imaging of the spatial distribution of particle numbers and to use ``wedding-cake'' structures as a possible signature for antiferromagnetism in the parameter regime corresponding to $U<U_{\rm c}$, where $U_{\rm c}$ is the critical interaction for Mott-insulator transitions. Furthermore, we found that the antiferromagnetic order parameter may always be chosen to be parallel to a globally defined direction and that there is a formation of domain-walls between checkerboard-type regions of the magnetization. The spins are aligned parallelly along these domain walls, which occur in regions where the filling factor changes from half-filling to higher or lower values.

\begin{acknowledgments}
We thank Sebastian Will (Univ.\ Mainz) for interesting discussions on the possible experimental observation of antiferromagnetic order as proposed in this paper.
\end{acknowledgments}

\bibliography{/home/gottwald/Masterbib/biblography.bib}

\end{document}